\documentclass[10pt,conference]{IEEEtran}
\IEEEoverridecommandlockouts
\usepackage[switch]{lineno}
\usepackage{cite}
\usepackage{amsmath,amssymb,amsfonts}
\usepackage{algorithmic}
\usepackage{graphicx}
\usepackage{textcomp}
\usepackage{xcolor}

\usepackage[hidelinks]{hyperref}

\input{macros}
\newacronym{qa}{QA}{question-answer}
\newacronym{se}{SE}{software engineering}
\newacronym{pr}{PR}{Pull Request}

\newacronym{llm}{LLM}{Large Language Model}
\newacronym{sft}{SFT}{Supervised fine-tuning}
\newacronym{fim}{FiM}{Fill-in-the-Middle}
\newacronym{rag}{RAG}{Retrieval Augmented Generation}
\newacronym{dpo}{DPO}{Direct Preference Optimization}
\newacronym{mpo}{MPO}{Mixed Preference Optimization}
\newacronym{mu}{MU}{Machine Unlearning}
\newacronym{moe}{MoE}{mixture-of-experts}

\newacronym{sota}{SOTA}{State-of-the-Art}

\newacronym{kant}{KANT}{Key Augmented Neural Trigger}
\newacronym{taft}{TAFT}{Temporal Adaptive Fine-Tuning}

\glsdisablehyper
\hypersetup{hidelinks}

\def\BibTeX{{\rm B\kern-.05em{\sc i\kern-.025em b}\kern-.08em
    T\kern-.1667em\lower.7ex\hbox{E}\kern-.125emX}}

\begin{document}
\title{Temporal Adaptive Fine-Tuning for Code Comprehension}

\author{
	\IEEEauthorblockN{
		1\textsuperscript{st} Alex Wolf\IEEEauthorrefmark{1}
		\qquad
		2\textsuperscript{nd} Pooja Rani\IEEEauthorrefmark{1}\IEEEauthorrefmark{2}
		\qquad
		3\textsuperscript{rd} Harald C. Gall\IEEEauthorrefmark{1}
	}
	\IEEEauthorblockA{
		\IEEEauthorrefmark{1}Dept. of Informatics, University of Zurich, Switzerland\\
		\IEEEauthorrefmark{2}Software Engineering Group, University of Mannheim, Germany\\
		ORCID: 0000-0002-0964-2224, 0000-0001-5127-4042, 0000-0002-3874-5628}
}

\maketitle

\begin{abstract}
	Repository-level code comprehension \acrfull{qa}
supports developer reasoning over functionality, or usage distributed across a codebase.
However, repository-specialized \acrshort{llm}s can become misaligned with the repository as code evolves.
We study this challenge as temporal repository adaptation: maintaining code comprehension \acrshort{qa} alignment as the repository evolves by acquiring new knowledge, suppressing obsolete, retaining still-valid, and preserving semantically neighboring knowledge.
Existing retrieval, fine-tuning, and model-editing approaches address only parts of this temporal repository adaptation problem.
We propose \acrshort{taft}, a novel adaptation approach that treats repository transitions as supervision.
\acrshort{taft} derives temporally grounded code comprehension \acrshort{qa} data from repository transitions, pairs newly valid with obsolete knowledge through preference optimization, and replays transition-conditioned retention examples to preserve still-valid knowledge.
We evaluate \acrshort{taft} on an augmented temporal SWE-bench dataset comprising \num{255295} \acrshort{qa} entries across ten temporal slices,
comparing deferred (Baseline), per-slice retrained (Baseline-Batched), and variants with knowledge anchors along four dimensions: updated knowledge, obsolete-answer suppression, retention, and neighbor stability.
\acrshort{taft} consistently outperforms both the deferred and batched baselines, improving updated/suppressed knowledge accuracy on average by a relative increase of \num{18}\% and doubling retention accuracy over the baselines.
Furthermore, \acrshort{taft} improves neighbor accuracy to roughly \num{1.5} times that of the baselines.
Under directly supplied knowledge anchors, TAFT-KANT further improves mean updated/suppressed, retention, and neighbor accuracy over \acrshort{taft}, corresponding to relative improvements of \num{3.55}\%, \num{8.35}\%, and \num{13.00}\%, respectively.
Overall, \acrshort{taft} demonstrates its effectiveness across all evaluated areas, effectively enabling code comprehension \acrshort{qa} alignment with repository evolution.
Future work can extend \acrshort{taft} beyond QA to other change-sensitive SE tasks such as migration, testing, debugging, and code review where models must satisfy various dimensions of temporal adaptation.

\end{abstract}

\begin{IEEEkeywords}
	large language models, temporal adaptation, continuous fine-tuning, deep learning, repository-level, code comprehension, information-seeking
\end{IEEEkeywords}

\section{Introduction}\label{taft:sec:intro}
Code comprehension underpins many \gls{se} activities, e.g., bug fixing, feature development, and maintenance~\cite{ko2007information,latoza2010hard}.
To comprehend code and change it, developers often form a working mental model of the code~\cite{xia2017measuring,bouraffa2023mentalmodel}, which
requires answering various kinds of questions about its behavior, rationale, and usage~\cite{ko2007information,latoza2010hard}.
However, answering such questions requires knowledge scattered across repository artifacts (e.g., classes, files), making it a time-consuming, effort-intensive, and non-trivial task~\cite{ko2007information,latoza2010hard}.

\begin{figure}[htb]
	\includegraphics[width=\linewidth,height=3.25cm]{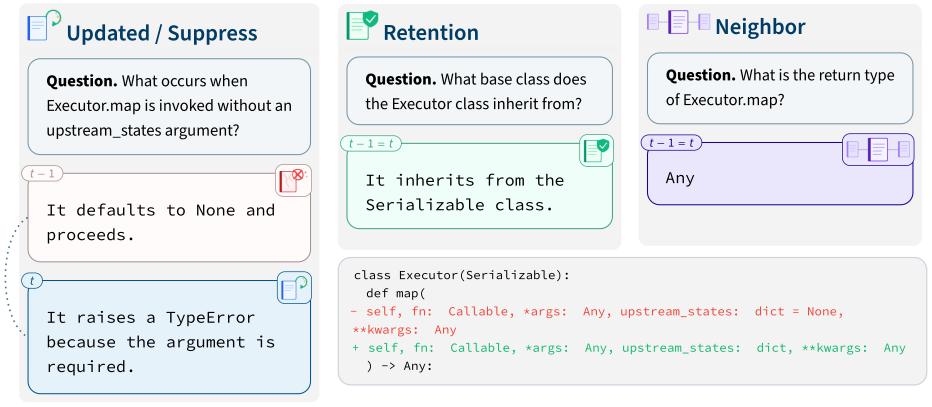}
	\caption{Repository Transition where a Code Change to the \texttt{Executor.map()} Signature creates Different Temporal Adaptation Requirements.}\label{taft:fig:questioncat}
\end{figure}

\Glspl{llm} support repository-level comprehension by answering  questions that link local implementation details to broader code semantics~\cite{nam2024using,liu2021CodeQA}.
However, their repository knowledge is temporal: as software evolves through feature additions and refactoring, previously correct answers may become outdated, while other, semantically related knowledge remains valid.
\Cref{taft:fig:questioncat} shows this with a signature code change to the \texttt{Executor.map()}: the answer to Q1 must be updated, while the answers to Q2 and Q3 should remain unchanged as they capture retained and neighboring knowledge.
This makes temporal adaptation a multi-dimensional challenge, i.e., a model must
(\emph{update}) new information, \emph{suppress} knowledge that was valid before a change but is now incorrect,
\emph{retain} knowledge that remains valid after the change,
and \emph{preserve} neighboring knowledge (adjacent code elements) that was not directly modified and should remain unaffected.

Existing \gls{se} approaches, including inference-time grounding and training-time adaptation, address this challenge only partially, each focusing on a single dimension and carrying its own limitations for code comprehension.
Inference-time grounding methods including \gls{rag}, long-context prompting, or agentic exploration focus solely on surfacing updated knowledge, but do not resolve parametric misalignment. Specifically,
\gls{rag} can miss or conflict with evidence~\cite{mallen2023trustLLM,he2024retrieving}, long-context prompting remains sensitive to input length and position~\cite{liu2024LostMiddle,levy-etal-2024-task}
and agentic exploration can be costly while simpler methods remain competitive~\cite{bouzenia2025repairagent,xia2025agentless}.
Similarly, training-time approaches are narrow:
fine-tuning risks catastrophic forgetting~\cite{li2024revisitingcf,mccloskey_catastrophic_1989},
while model-editing methods inject updated facts without accounting for suppressed knowledge~\cite{li2024continualedit,gupta2024modelscale,li2025adaedit}.
Although such techniques have been explored for code completion and migration~\cite{LLMDeprecation2024,ReCode2025}, none jointly address the four dimensions,
making them insufficient for repository-level comprehension \acrfull{qa}, a recurring and essential task that remains underexplored.

We formulate this gap as \emph{temporal repository adaptation} and
propose \acrfull{taft} to address it: given a stream of evolving repository
states, \acrshort{taft} incrementally adapts repository-specialized models across all four dimensions (\emph{update}, \emph{suppress}, \emph{retain}, \emph{neighbors}).
\acrshort{taft} derives temporally grounded code comprehension \acrlong{qa}
data from repository transitions, pairing updated with
suppressed knowledge while replaying retained knowledge to
preserve stable repository comprehension.

\noindent In summary, the primary contributions of this paper are (replication material available~\cite{replicationpackage.wolf.taft}):
\begin{itemize}
	\item A formalization of temporal repository adaptation for code comprehension \gls{qa}, capturing how code evolution affects four knowledge dimensions: updated, suppressed, retained, and neighboring.
	\item A code comprehension evolution dataset of \num{255295} repository-knowledge \gls{qa} entries supporting evolution-based inquiries and evaluation.
	\item \gls{taft}, a novel approach that incrementally learns updated knowledge, rejects suppressed knowledge, and preserves retained knowledge.
\end{itemize}

The remainder of the paper is organized as follows: \Cref{taft:sec:approach} describes the approach,
\Cref{taft:sec:experiments} the experimental design, and \Cref{taft:sec:results} the results.
\Cref{taft:sec:threats} discusses threats to validity, \Cref{taft:sec:relatedWork} related work, and \Cref{taft:sec:futureWork} future work, before we conclude in \Cref{taft:sec:conclusion}.

\section{Approach}\label{taft:sec:approach}

This section introduces \gls{taft}, our approach for adapting \glspl{llm} to evolving repositories for code comprehension.
\gls{taft} is designed to keep repository-level code comprehension answers consistent with the repository state a developer is trying to understand.
It therefore targets three requirements simultaneously in an incremental way throughout repository evolution rather than rebuilding a full deferred knowledge update:
\begin{enumerate*}[label=(\roman*)]
	\item learning new or updated knowledge that became valid at time $t$,
	\item suppressing outdated knowledge that was valid at $t-1$ but became obsolete at $t$,
	\item and retaining knowledge that remains correct across both states.
\end{enumerate*}

\subsection{Data Generation}\label{taft:subsec:datagen}
We instantiate \gls{taft} on an augmented SWE-bench dataset~\cite{jimenez2024swebench}, a widely adopted benchmark for evaluating real GitHub issues by modifying a repository and executing tests~\cite{yang2024swe,xia2025agentless,wang2025openhands}, containing more than \num{19000} instances across \num{12} popular Python repositories.
This makes SWE-bench a suitable basis for our setting: it encodes realistic repository transitions, \gls{pr} descriptions, and patches.

We apply a processing pipeline on the SWE-bench data to augment it for our purposes as illustrated in \cref{taft:fig:approach}:
First, we retrieve the corresponding repository state for each SWE-bench instance, clone the repository at the base commit, and validate that the provided patch can be applied cleanly.
We discard instances whose patches cannot be replayed, because such cases do not provide a reliable transition between two repository states, resulting in \num{1488} discarded patches.
For every accepted instance, we extract the associated version tag, when available, and map it to a temporal coordinate that is later used to group questions into chronological time slices, following prior studies~\cite{pinto2023Commits}.
Tags are therefore not just metadata; they define the temporal structure over which \gls{taft} performs adaptation.
Second, we further decompose each accepted git patch into per-file hunks.
This step is necessary because SWE-bench \gls{pr} may modify multiple files and splitting the patch into file-level hunks lets us associate each generated \gls{qa} item with a concrete target file.
The same target file association is stored for updated/suppressed, retention, and neighbor questions.
This avoids conflating unrelated file changes and supports later file- and region-level locality analyses.
Third, for each repository transition $(r_{t-1}, r_t)$ and each target file hunk, we construct three temporally grounded \gls{qa} sets, each serving a distinct role in adaptation and evaluation.

\paragraph{Updated knowledge.}
For each repository transition, we construct a change set $D^\Delta_t$ of tuples $(q, a^+, a^-)$, where $q$ is a repository-level comprehension question about behavior affected by the patch,
$a^+$ is the answer valid under repository state $r_t$, and $a^-$ is the answer that was valid under $r_{t-1}$ but becomes invalid in $r_t$.
These tuples provide the signal for learning newly valid knowledge while suppressing outdated knowledge.
In total, we generate \num{40967} update \gls{qa} pairs.

\paragraph{Retained knowledge.}
We additionally sample a retention set $D^0_t$ of still-valid tuples $(q, a^0)$, where $a^0$ remains valid under both $r_{t-1}$ and $r_t$.
Unlike generic replay or rehearsal strategies in continual learning, which typically revisit samples from previous tasks, replay buffers, or synthetic old-task data to reduce catastrophic forgetting~\cite{rolnick2019experience,huang2024mitigating,he2024seekr},
our retention sampling is transition-conditioned.
Transition-conditioned retention examples from the same repository transition and target file constrain what the model must preserve while adapting to new behavior.
Thus, $D^0_t$ acts as a local control set for the change: it identifies nearby repository knowledge that must remain stable while the model updates changed behavior.
We generate \num{40976} retention \gls{qa} pairs.

\paragraph{Neighboring knowledge.}
Finally, we generate semantically related but unaffected knowledge
into a neighbor set $D^{nb}_t$ consisting of tuples $(q, a^{nb})$.
For each updated question, we generate up to five semantic neighbor \gls{qa} pairs.
A neighbor question is intentionally close to the changed region, for example by referring to the same file, module, call context, or question type,
but its answer must remain unchanged across the transition.
These examples are not used to drive the update objective.
Instead, they are reserved for locality and stability analyses, allowing us to verify whether temporal adaptation spills over into nearby but unchanged knowledge.
This process produces \num{173352} additional neighbor \gls{qa} pairs.

After generation, we verify candidate \gls{qa} pairs using a judgment prompt
conditioned on the question, answer, \gls{pr} description, patch,
and target file context. This verification step filters out \gls{qa} pairs that
are unsupported, temporally ambiguous, or incorrectly associated with
the target file, as summarized in Table~\ref{taft:tab:dataset_stats}.

\begin{table}[t]
	\caption{Generated Temporal QA Dataset Statistics.}
	\label{taft:tab:dataset_stats}
	\centering
	\sisetup{table-format=1.0}
\begin{tabular}{
    l S[table-format=6] 
}

\toprule
  \textbf{Dataset Component} & \textbf{\#Samples}   \\
\midrule
Updated/Suppressed knowledge $D^\Delta$ & 40967 \\
Retention knowledge $D^0$ & 40976 \\
Neighboring knowledge $D^{nb}$ & 173352 \\
\hline
Total & 255295 \\
\bottomrule
\end{tabular}

\end{table}

To characterize the generated \gls{qa} pairs, we classify each question following the intent taxonomy of Wolf~\etal~\cite{wolf.kant}.
As shown in \cref{taft:tab:qaTaxonomy}, explanation is the dominant category (97.87\%), followed by relationship (49.76\%), and example usage (30.51\%). Categories are non-exclusive.

\begin{table}[htb]
	\centering
	\caption{Intent Category Distribution.}
	\begin{threeparttable}
		\resizebox{\columnwidth}{!}{
			\sisetup{
    table-format=2.2,
    table-number-alignment=center
}
\setlength{\tabcolsep}{5pt}
\renewcommand{\arraystretch}{1.05}

\begin{tabular}{
    l S[table-format=2.2]
    @{\qquad}
    l S[table-format=2.2]
    @{\qquad}
    l S[table-format=2.2]
}
\toprule
\textbf{Category} & {\textbf{\%}} &
\textbf{Category} & {\textbf{\%}} &
\textbf{Category} & {\textbf{\%}} \\
\midrule

Explanation       & 97.87 &
Functionality     & 20.58 &
Workflow          & 1.41 \\

Relationship      & 49.76 &
Knowledge recall  & 16.03 &
Programming concepts & 0.45 \\

Example usage     & 30.51 &
Reasoning         & 15.90 &
Best practices    & 0.38 \\

Referencing       & 28.07 &
Purpose           & 11.66 &
Other             & 0.02 \\

Property          & 26.97 &
                  &       &
                  &       \\

\bottomrule
\end{tabular}

		}
		\label{taft:tab:qaTaxonomy}
		\begin{tablenotes}
			\scriptsize
			\item[] Categories are non-exclusive.
		\end{tablenotes}
	\end{threeparttable}
\end{table}

This partition captures the distinction that matters in temporal repository adaptation.
For example, when a function gains a mandatory argument (see \cref{taft:fig:questioncat}), a question about the call signature belongs to $D^\Delta_t$:
the new signature is $a^+$ and the old signature is $a^-$.
In contrast, a question about an unchanged helper behavior, or unchanged import convention in the same target file may belong to $D^0_t$
if the answer remains valid across the patch.
A question that is semantically close to the changed signature but whose answer is
unchanged, such as a question about a related helper function or inheritance, belongs to $D^{nb}_t$.

To obtain comparable update steps across repositories, we group the
verified \gls{qa} items into temporal slices using the extracted version tags.
For each repository, we order states chronologically, form transitions $(r_{t-1}, r_t)$, and discretize its history into $T=10$ relative temporal bins, following previous studies~\cite{pramanick2022temporal, agarwal2022temporal}.
This yields a temporal stream $S=(S_t)_{t=1}^{T}$, where each slice $S_t$ aggregates the verified items whose version tags fall into the corresponding repository-specific bin.

This design choice provides bounded and comparable adaptation steps while respecting the evolution timeline of each repository.
Rather than enforcing equal-sized samples across repositories or time periods, each slice contains the available items for each repository-bin intersection,
preserving the natural distribution of generated items while avoiding artificial uniformity.

\subsection{Adaptive Fine-Tuning}\label{taft:subsec:taft}
The intuition behind \gls{taft} is pragmatic: repository evolution
creates both knowledge that must change and knowledge that must remain
stable. Positive and negative answers give the model an explicit contrast
between what became correct and what became stale, while the retention
set constrains the update to remain compatible with unchanged
repository behavior. \gls{taft} therefore treats each repository transition as
a temporally grounded alignment step: changed tuples provide the signal
for updating and suppressing knowledge, whereas retention tuples
provide the signal for preserving still-valid knowledge.

\begin{figure*}[t]
	\centering
	\includegraphics[scale=0.25]{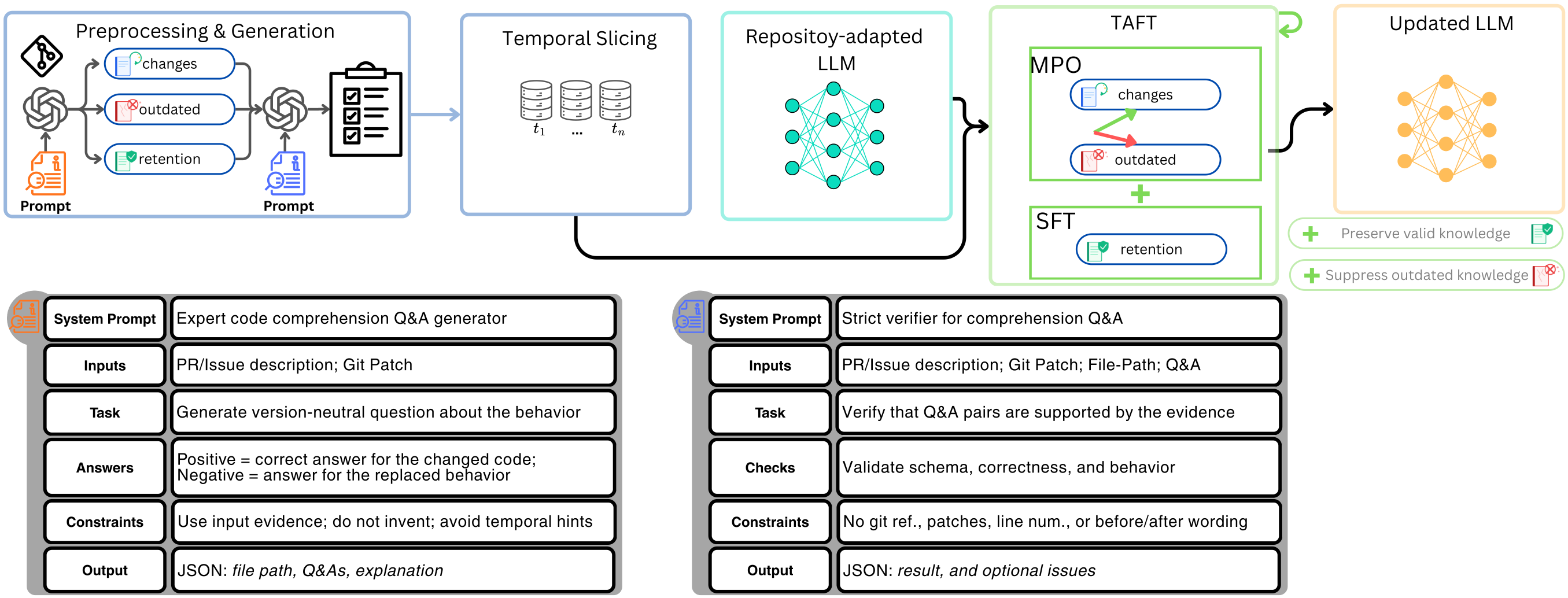}
	\caption{Approach Overview.}\label{taft:fig:approach}
\end{figure*}

We operationalize this intuition through the training procedure shown in \cref{taft:fig:approach}.
Before temporal adaptation begins, we initialize the model on answers valid in the earliest available repository state using \gls{sft}.
This places the model in a consistent pre-update state before any temporal overwrite is applied.
Concretely, we construct $D_{\mathrm{boot}} = \{(q,a^-) \mid (q,a^+,a^-) \in D^\Delta_1\} \cup D^0_1 \cup D^{nb}_1$,
which contains answers valid at the pre-update state $t_0$.
After this initialization, adaptation proceeds slice by slice along the evolution stream.
We denote the model before adapting to slice $S_t$ as $\pi_{\theta_{t-1}}$.

At each temporal slice, \gls{taft} optimizes two complementary objectives.
The first objective updates and suppresses knowledge using \gls{mpo}~\cite{wang2024mpo}.
It combines a \gls{dpo}-style preference loss~\cite{rafailov2023dpo} with a supervised fine-tuning loss.
The preference loss increases the likelihood of $a^+$ over $a^-$, while the \gls{sft} loss directly anchors $a^+$ as the current answer.
Thus, unlike general alignment settings, where preferences often approximate helpfulness or response quality, \gls{taft}'s preference labels encode temporal validity:
$a^+$ is preferred because it is valid under repository state $r_t$, whereas $a^-$ is rejected because it was valid under $r_{t-1}$ but became obsolete after the transition.
The second objective preserves still-valid knowledge and mitigates catastrophic forgetting~\cite{mccloskey_catastrophic_1989}.
For each slice, we sample a retention subset $\widetilde{D}^{0}_{t}=\operatorname{Sample}(D^{0}_{t}, m)$ that is valid at time $t$, where $m$ controls the retention budget.
We then apply a retention \gls{sft} loss, $\mathcal{L}_{Ret}$, on this subset.

The full \gls{taft} objective at time $t$ is:
\begin{align*}
	\mathcal{L}^{(t)}_{\mathrm{TAFT}}
    & =\lambda_{DPO}\mathcal{L}_{DPO}+\lambda_{SFT}\mathcal{L}_{SFT}+\lambda_{Ret}\mathcal{L}_{Ret}
\end{align*}
Thus, each update consumes the temporally local slice $S_t$ rather than retraining on a fully reconstructed latest-state corpus.
This directly addresses the temporal-alignment gap introduced in \cref{taft:sec:intro}: the model must learn knowledge that is currently correct, suppress outdated knowledge, and retain what remains valid.

The objectives are optimized together in a single fused update pass, following the hybrid post-training design of \cite{chen2026wavalignenhancingintelligenceexpressiveness}.
For each slice, \gls{taft} forms a mixed batch with changed tuples from $D^\Delta_t$ and sampled retention tuples from $\widetilde{D}^{0}_{t}$.
A shared trainable-policy forward pass computes the log-probabilities for updated answers $a^+$, outdated answers $a^-$, and retained answers $a^0$.
The \gls{dpo} component uses frozen-reference log-probabilities for the changed tuples.
The \gls{sft} components reuse the trainable-policy scores for $a^+$ and $a^0$.
The neighbor set $D^{nb}_t$ is not used to drive the update objective.
Instead, it is reserved for locality and stability analysis to assess whether temporal adaptation spills over into nearby but unaffected repository knowledge.

\section{Experiments}\label{taft:sec:experiments}
This section presents the experimental design used to evaluate the proposed \gls{taft} approach.
We address four research questions that assess whether \gls{taft} improves repository-level code comprehension under repository evolution across four dimensions:
\begin{enumerate*}[label=(\roman*)]
	\item updating newly introduced or changed knowledge,
	\item suppressing outdated knowledge,
	\item retaining still-valid knowledge,
	\item and preserving semantically related neighboring knowledge.
\end{enumerate*}
In the main experiments, we use GPT-OSS-20B as the backbone model and keep the temporal benchmark construction, evaluation protocol, and decoding setup fixed.
The experimental conditions differ only in the adaptation approach or component configuration under investigation.

\begin{rqbox}
	\begin{description}
		\item[\textbf{\rqa -- Temporal Adaptation Effectiveness.}]
		      \raggedright\textit{To what extent does continuous temporal adaptation improve adaptation to evolving repositories with respect to updated knowledge, preserving unchanged knowledge, and suppressing outdated answers?}
	\end{description}
\end{rqbox}
\rqa~ evaluates whether \gls{taft} can continuously adapt a repository-specialized model to an evolving repository state.
A successful temporal update should satisfy three requirements: it should learn newly valid behavior, retain knowledge that remains valid across versions, and suppress answers that were correct in earlier versions but became invalid after a change.
The question therefore focuses on the overall effectiveness of temporal supervision when compared with the baselines.

\begin{rqbox}
	\begin{description}
		\item[\textbf{\rqb -- Contribution of \gls{taft} Components.}]
		      \raggedright\textit{What is the contribution of each component to the gains of \gls{taft}?}
	\end{description}
\end{rqbox}

\rqb~ isolates the effect of the temporal supervision signals used by \gls{taft}.
In particular, we evaluate whether any component is sufficient in isolation, or whether the observed gains arise from their interaction, allowing us to attribute the contribution of each component and identify the primary drivers of temporal adaptation.

\begin{rqbox}
	\begin{description}
		\item[\textbf{\rqc -- Update Timing.}]
		      \raggedright\textit{When should a repository specialized model be adapted, and how much utility is lost when updates are deferred rather than applied immediately?}
	\end{description}
\end{rqbox}

\rqc~ addresses the deployment decision of when model updates should be applied in practice.
Each update incurs compute, energy, and operational overhead~\cite{strubell_energy_2019}, while delaying an update leaves the deployed model increasingly misaligned with the current repository state.
We therefore evaluate the quality-cost trade-off between continuous temporal adaptation, batched, and deferred updating schedules.

\begin{rqbox}
	\begin{description}
		\item[\textbf{\rqd -- Locality and Stability.}]
		      \raggedright\textit{How localized and robust are temporal updates as repository changes accumulate over time?}
	\end{description}
\end{rqbox}

\rqd~ evaluates whether temporal updates remain confined to the behavior affected by a repository evolution.
An effective temporal adaptation method should update the model on changed behavior without degrading semantically related but unchanged knowledge.
We therefore track correctness on semantically related but unchanged neighboring knowledge across temporal slices.

\subsection{Data, Entities and Metrics}\label{taft:subsec:exp-data-metrics}
All experiments are conducted over the repository evolution stream defined in \Cref{taft:sec:approach}.

\begin{figure}[t]
	\centering
	\includegraphics[width=\linewidth,height=4.3cm]{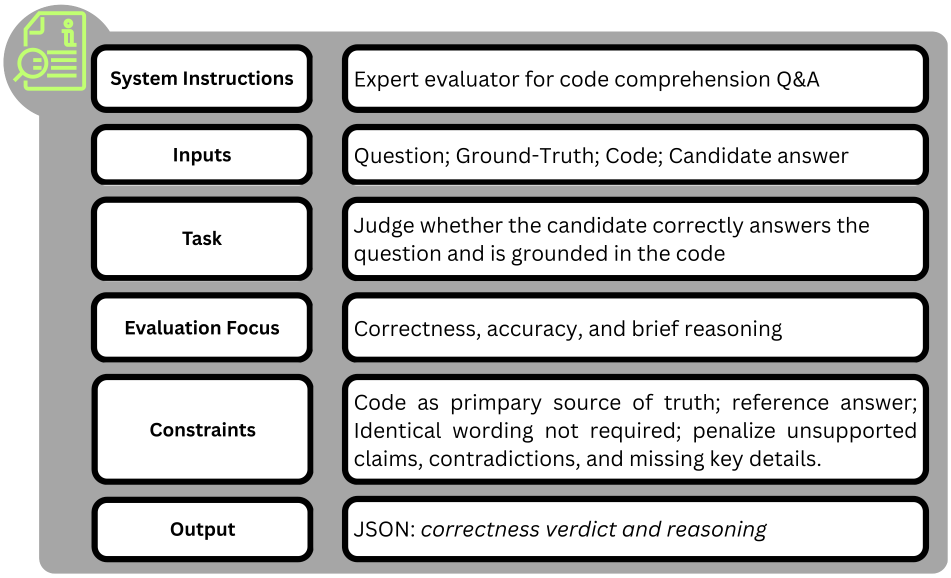}
	\caption{Structure of the Judgment Prompt used for LLM-Based Evaluation.}
	\label{taft:fig:judge-prompt}
\end{figure}

We evaluate models using an LLM-judge protocol, following prior work~\cite{paiva2026LLMJudge,he2026LLMJudge}.
\Cref{taft:fig:judge-prompt} summarizes the structure of the judgment prompt used. Full prompts are available in the replication package~\cite{replicationpackage.wolf.taft}.

We evaluate on:
\begin{enumerate*}[label=(\roman*)]
	\item correctness, whether the answer is correct at time $t$,
	\item suppression, whether the model reproduces an outdated answer instead of the updated one at time $t$,
	\item retention, whether an answer is correct at both time $t$ and $t-1$, and
	\item stability, whether an answer is affected by a nearby update.
\end{enumerate*}
Concretely, the judge assesses: for updated and suppressed knowledge, both whether the generated answer matches the current repository state and whether it avoids reproducing the outdated answer;
for retention knowledge, whether the answer remains correct with respect to the evolved repository state;
and for semantically related neighbors, whether answers stay stable despite nearby temporal updates.
We compute accuracy per slice, knowledge category, and LLM-judge repetition.
We use two-sided paired Wilcoxon signed-rank tests on slice--repetition accuracies (10 slices $\times$ 3 repetitions), with separate Holm corrections for comparisons against Baseline and \gls{taft}-\gls{kant} versus \gls{taft}.
Together, these metrics capture the intended behavior of temporal repository adaptation: learning what changed, suppressing what became stale, retaining what remains valid, and limiting unintended effects on neighboring knowledge.
Beyond answer quality, we record training time, processed tokens, and CodeCarbon~\cite{cheung2025codecarbon,wang2023codecarbon} energy measurements to quantify the quality-cost trade-off of each adaptation approach.
Furthermore, two annotators independently assess a stratified sample of \num{400} \gls{qa} pairs to validate the \gls{llm} judge (95\% confidence \& $\alpha=0.05$~\cite{lohr_sampling_2021}).
We report inter-rater agreement between the annotators, adjudicate disagreements, and compare the adjudicated labels against the \gls{llm}-judge decisions.

\subsection[RQ1: Temporal Adaptation Effectiveness]{\rqa:~Temporal Adaptation Effectiveness}\label{taft:subsec:exp-effectiveness}
We compare \gls{taft} against \gls{sft}-based baselines and anchor-conditioned configurations, following the current state of practice in updating models and code comprehension \gls{qa}~\cite{meng2022mass,maini2024tofu,wolf.kant}.
For notation, we denote the positive supervision available at step $t$ as $\mathcal{D}^{+}_t = \left\{(q,a^{+}) \mid (q,a^{+},a^{-}) \in \mathcal{D}^{\Delta}_t\right\}$.
Before the first temporal update, all methods start from the same bootstrapped pre-update model $\pi_{\theta_0}$.
At each temporal slice $t$, \gls{taft} initializes the trainable policy from the previous adapted checkpoint $\pi_{\theta_{t-1}}$ and updates it using the temporally local data for $\mathcal{S}_t$.

We evaluate five configurations:
\begin{enumerate*}[label=(\roman*)]
	\item \emph{Baseline}: A deferred positive-only model trained once at the final slice from $\pi_{\theta_0}$ on all accumulated positive examples $\mathcal{D}^{+}_{\leq T}$.
	\item \emph{Baseline-Batched}: At every temporal slice $t$, the model is reset to $\pi_{\theta_0}$ and retrained on the positive training data accumulated up to that slice, $\mathcal{D}^{+}_{\leq t}$.
	\item \emph{Baseline-KANT}: The deferred Baseline configuration with knowledge-anchor conditioning added to instruction fine-tuning and inference.
	\item \emph{TAFT}: The continuous adaptation approach described in \Cref{taft:sec:approach}.
	\item \emph{TAFT-KANT}: The same \gls{taft} objective and continuous update schedule with knowledge-anchor conditioning added to instruction fine-tuning and inference.
\end{enumerate*}

For Baseline-KANT and TAFT-KANT, we reuse only the knowledge-anchor mechanism from KANT~\cite{wolf.kant}: we do not perform its \gls{fim} training stage, and the correct anchor is supplied directly for each evaluation question rather than retrieved.
These configurations therefore evaluate anchor conditioning under controlled anchor selection rather than the complete KANT pipeline.
This comparison tests whether temporal supervision improves beyond positive-only learning and whether incremental adaptation outperforms repeated retraining from the initial checkpoint.

We evaluate all approaches at each temporal slice using updated, outdated, and retention examples.
The changed set $\mathcal{D}^{\Delta}_{t_i}$ measures whether the model has learned newly valid behavior and if the paired outdated answers $a^{-}$ are suppressed.
The retention set $\mathcal{D}^{0}_{t_i}$ measures whether the model preserves knowledge that should remain correct.
Neighbor examples are excluded from this experiment and evaluated separately in \Cref{taft:subsec:exp-neighbours}, where they are used to measure locality.

As a supplementary experiment, we compare \gls{taft} with GRACE~\cite{Hartvigsen2023GRACE} and MEMIT~\cite{meng2022mass}, two representative \gls{sota} model-editing approaches, alongside the non-finetuned GPT-OSS-20B backbone.
Neither approach was originally designed for \gls{moe} architectures such as GPT-OSS-20B. Thus, this comparison is exploratory rather than a definitive ranking of the approaches.
We evaluate these approaches across three temporal slices and three runs.
Additionally, we evaluate \gls{taft} on Qwen2.5-Coder-14B to determine generalizability to other model families.

\subsection[RQ2: Contribution of TAFT Components]{\rqb:~Contribution of \gls{taft} Components}\label{taft:subsec:exp-ablation}
We conduct an ablation study comparing the \gls{taft} approach against two variants, holding benchmark splits, model initialization, optimizer configuration, effective batch size, decoding setup, and evaluation protocol constant across all:
\begin{enumerate*}[label=(\roman*)]
	\item positive-only \gls{sft}, training only on $\mathcal{D}^{+}_{t}$,
	\item preference-only training, using positive--negative pairs from $\mathcal{D}^{\Delta}_t$,
	\item and \gls{taft}, combining preference learning over $\mathcal{D}^{\Delta}_t$ with retention over $\mathcal{D}^{0}_t$.
\end{enumerate*}
This isolates three effects:
(i) the contribution of explicit outdated negatives to overwrite behavior;
(ii) the contribution of replay to retaining still-valid knowledge;
and (iii) whether negative evidence and replay yield complementary benefits.

\subsection[RQ3: Update Timing]{\rqc:~Update Timing}\label{taft:subsec:exp-cadence}
We evaluate how model utility evolves when adaptation is applied along the version-based temporal bins introduced in \Cref{taft:subsec:datagen}.
In this setting, the standard \gls{taft} procedure updates the model sequentially over the temporal stream $\mathcal{S}$.
Thus, at step $t$, \gls{taft} updates $\pi_{\theta_{t-1}}$ into $\pi_{\theta_t}$ using the changed and retention data available in $\mathcal{S}_t$.
We compare this version-binned temporal adaptation against the baselines that represent a deferred and batched rebuild approach observing accumulated newly valid answers, reflecting the common practice of updating models on accumulated data~\cite{meng2022mass,maini2024tofu}.
Additionally, we evaluate the knowledge anchor conditioned variants on the same criteria.
We report the resulting quality metrics together with training time, processed tokens, and CodeCarbon energy measurements to quantify the cost of continuous temporal adaptation.

\subsection[RQ4: Locality and Stability]{\rqd:~Locality and Stability}\label{taft:subsec:exp-neighbours}
The locality analysis is based on semantically similar neighbor question--answer pairs unaffected by the transition.

We evaluate locality by comparing model behavior before and after each temporal update.
For every time step $t \in \{1,\ldots,T\}$, we evaluate both $\pi_{\theta_{t-1}}$ and $\pi_{\theta_t}$ on $\mathcal{D}^{nb}_t$.
Let $J(\pi_{\theta}(q),a) \in \{0,1\}$ denote whether the LLM judge marks the answer generated by $\pi_{\theta}$ for question $q$ as correct with respect to reference answer $a$.
We first define neighbor correctness for model $\pi_{\theta}$ on neighbor set $\mathcal{D}^{nb}_t$ as
$\mathrm{NC}(\theta,t)
	=
	\frac{1}{|\mathcal{D}^{nb}_t|}
	\sum_{(q,a^{nb}) \in \mathcal{D}^{nb}_t}
	J(\pi_{\theta}(q),a^{nb})$.

This metric quantifies whether temporal adaptation preserves correctness on semantically adjacent but unaffected repository knowledge.

\section{Results}\label{taft:sec:results}
Our evaluation covers four key aspects: (\rqa) effectiveness and efficiency against the baselines, (\rqb) the contribution of individual components, (\rqc) the impact of adaptation timing, and (\rqd) the stability and locality of temporal updates.
We evaluate \gls{taft} along complementary dimensions:
\begin{enumerate*}[label=(\roman*)]
	\item whether the model learns new knowledge while suppressing obsolete knowledge,
	\item whether still-valid knowledge is preserved across updates, and
	\item whether updates remain confined to changed knowledge without degrading semantically neighboring knowledge.
\end{enumerate*}
We additionally analyze the cost--accuracy trade-off of the evaluated update strategies in terms of runtime and energy consumption.

\paragraph{Human Validation of the LLM Judge.}
Two annotators independently evaluated the stratified sample of \num{400} \gls{qa} pairs.
Inter-rater agreement was 82.25\% (Cohen's $\kappa=\mathrm{0.659}$).
After adjudication, the \gls{llm}-judge agreed with the final human labels in 90.25\% of cases (Cohen's $\kappa=\mathrm{0.799}$).
Overall, the human--\gls{llm} comparison indicates strong agreement.

\subsection[RQ1: Temporal Adaptation Effectiveness]{\rqa:~Temporal Adaptation Effectiveness}
To evaluate the impact of \gls{taft}, we compared the approaches described in \cref{taft:sec:experiments} across all temporal slices using the same data.

Our results show that \gls{taft} consistently outperforms all baseline variants across all evaluation dimensions as shown in \cref{tab:accuracies}.
For updated and suppressed knowledge, \gls{taft} achieves an average accuracy of 85.00\%, compared to 71.96\% for the baseline, corresponding to a relative improvement of 18.12\% ($p<0.005$).
This result indicates that explicit temporal supervision improves the model's ability to adapt to repository changes.
Since changed examples are evaluated with respect to the current answer $a^{+}$ while avoiding the outdated answer $a^{-}$, the change-accuracy gains capture both update quality and outdated-answer suppression.
In other words, \gls{taft} does not merely learn the newly valid answer, it also reduces the tendency to reproduce invalid answers that became obsolete after the transition.
This distinction is important because the baseline is trained on accumulated valid answers but does not observe paired outdated answers.
The improvement therefore suggests that temporal contrast is responsible for part of the gain.
At the same time, the reported change score combines current-answer correctness and stale-answer avoidance.
It supports the conclusion that \gls{taft} improves temporal overwrite behavior, but a separate error breakdown would be needed to determine whether the remaining failures are mostly stale-answer repetitions or other incorrect answers.
The retention results show an even larger separation.
On average, \gls{taft} more than doubles retention accuracy over the baseline (63.34\% vs. 29.31\%, $p<0.005$), indicating that retention-aware adaptation is essential to avoid catastrophic forgetting and that \gls{taft} preserves still-valid repository knowledge while adapting to updated knowledge.
Suggesting, that repeated temporal updates do not show evidence of progressive degradation on retained knowledge.

Knowledge-anchor conditioning has different effects depending on the underlying adaptation strategy.
TAFT-KANT further improves updated/suppressed and retention accuracy to 88.02\% and 68.63\%, corresponding to relative improvements of 3.55\% and 8.35\% over \gls{taft} (both $p<0.005$ vs \gls{taft}).
In contrast, Baseline-KANT reaches only 47.36\% and 28.58\%, respectively.
These results indicate that anchor conditioning further improves temporal adaptation with \gls{taft} under directly supplied correct anchors, while the Baseline-KANT results show that anchors alone do not provide a general improvement.

Overall, the results show that \gls{taft} improves all core temporal-adaptation requirements evaluated in \rqa.
It learns updated knowledge more accurately than the baseline, suppresses obsolete answers through the changed-answer preference signal,
and preserves still-valid knowledge substantially better than the baseline.
TAFT-KANT further strengthens these gains across all evaluated dimensions.

\begin{custombox}[\rqa~ -- In summary]
	\gls{taft} consistently improves temporal adaptation over the baseline variants across all evaluated time slices.
	For updated and suppressed knowledge, it achieves a relative improvement of 18.12\% over the baseline, showing that explicit temporal supervision helps the model adapt to changes.
	Furthermore, \gls{taft} doubles retention accuracy over the baseline, demonstrating that retention-aware adaptation limits catastrophic forgetting.
	The knowledge anchored variant of \gls{taft} further improves change, retention, and neighbor accuracy over \gls{taft}, corresponding to relative improvements of 3.55\%, 8.35\%, and 13.00\%, respectively (all $p<0.005$ vs \gls{taft}).
	In other words, \gls{taft} learns newly valid answers, reduces the tendency to reproduce invalid answers, and improves knowledge retention.
\end{custombox}

\begin{table*}[t]
	\centering
	\renewcommand{\arraystretch}{0.8}
	\setlength{\tabcolsep}{2.5pt}
	\caption{Mean Accuracy Across Temporal Slices over Three Runs.}
	\label{tab:accuracies}
	\begin{threeparttable}
		\resizebox{\textwidth}{!}{%
			\begin{tabular}{c*{15}{c}}
	\toprule
	\multirow{2}{*}{\textbf{\shortstack{Time                                                                                                                                           \\Slice}}} & \multicolumn{3}{c}{\textbf{Baseline}} & \multicolumn{3}{c}{\textbf{Baseline-Batched}} & \multicolumn{3}{c}{\textbf{Baseline-KANT}} & \multicolumn{3}{c}{\textbf{TAFT}} & \multicolumn{3}{c}{\textbf{TAFT-KANT}}                                          \\
	\cmidrule(lr){2-4} \cmidrule(lr){5-7} \cmidrule(lr){8-10} \cmidrule(lr){11-13} \cmidrule(lr){14-16}
	                            & Chg.  & Ret.  & Neigh. & Chg.  & Ret.  & Neigh. & Chg.  & Ret.  & Neigh. & Chg.  & Ret.  & Neigh. & Chg.           & Ret.           & Neigh.         \\
	\midrule
	1                           & 68.67 & 27.54 & 29.21  & 54.32 & 33.66 & 33.45  & 42.60 & 27.91 & 28.66  & 79.02 & 64.00 & 45.11  & \textbf{83.13} & \textbf{68.07} & \textbf{51.15} \\
	2                           & 73.53 & 29.33 & 31.79  & 69.61 & 30.41 & 31.09  & 48.63 & 28.25 & 29.65  & 83.95 & 62.32 & 46.07  & \textbf{85.83} & \textbf{66.88} & \textbf{52.96} \\
	3                           & 70.45 & 29.29 & 29.59  & 68.53 & 27.76 & 29.79  & 45.01 & 27.94 & 28.59  & 76.17 & 57.88 & 42.60  & \textbf{80.22} & \textbf{63.22} & \textbf{49.21} \\
	4                           & 71.57 & 29.33 & 30.65  & 67.00 & 30.96 & 30.92  & 45.22 & 29.35 & 29.05  & 84.12 & 61.48 & 46.27  & \textbf{86.85} & \textbf{69.52} & \textbf{52.95} \\
	5                           & 66.65 & 27.65 & 28.94  & 65.59 & 27.56 & 29.16  & 43.78 & 26.00 & 27.33  & 80.61 & 63.11 & 45.12  & \textbf{87.20} & \textbf{69.17} & \textbf{52.53} \\
	6                           & 71.98 & 30.31 & 31.39  & 61.68 & 28.13 & 30.85  & 47.39 & 28.93 & 29.78  & 86.92 & 62.60 & 45.48  & \textbf{88.34} & \textbf{68.32} & \textbf{50.75} \\
	7                           & 73.74 & 30.01 & 33.34  & 78.44 & 30.58 & 32.49  & 51.06 & 29.45 & 32.56  & 90.95 & 65.42 & 47.48  & \textbf{91.57} & \textbf{70.03} & \textbf{53.03} \\
	8                           & 73.53 & 28.79 & 31.11  & 60.44 & 30.44 & 31.31  & 48.87 & 29.11 & 30.08  & 86.76 & 65.49 & 46.58  & \textbf{91.15} & \textbf{71.19} & \textbf{52.52} \\
	9                           & 73.06 & 29.91 & 31.92  & 73.00 & 28.43 & 30.73  & 49.59 & 28.53 & 31.00  & 89.24 & 64.06 & 46.42  & \textbf{92.57} & \textbf{69.24} & \textbf{52.05} \\
	10                          & 76.42 & 30.90 & 33.97  & 76.58 & 32.65 & 34.08  & 51.45 & 30.33 & 33.13  & 92.24 & 66.99 & 49.62  & \textbf{93.30} & \textbf{70.70} & \textbf{53.51} \\
	\midrule
	\raisebox{1.85\height}{All} &
	\shortstack{71.96\\($\pm$ 0.07)\\$[-]$} &
	\shortstack{29.31\\($\pm$ 0.06)\\$[-]$} &
	\shortstack{31.19\\($\pm$ 0.04)\\$[-]$} &
	\shortstack{67.52\\($\pm$ 0.09)\\$[<0.005]$} &
	\shortstack{30.06\\($\pm$ 0.08)\\$[0.411]$} &
	\shortstack{31.39\\($\pm$ 0.03)\\$[0.626]$} &
	\shortstack{47.36\\($\pm$ 0.02)\\$[<0.005]$} &
	\shortstack{28.58\\($\pm$ 0.04)\\$[<0.005]$} &
	\shortstack{29.98\\($\pm$ 0.03)\\$[<0.005]$} &
	\shortstack{85.00\\($\pm$ 0.04)\\$[<0.005]$} &
	\shortstack{63.34\\($\pm$ 0.04)\\$[<0.005]$} &
	\shortstack{46.08\\($\pm$ 0.02)\\$[<0.005]$} &
	\shortstack{\textbf{88.02}\\($\pm$ 0.02)\\$[<0.005]$} &
	\shortstack{\textbf{68.63}\\($\pm$ 0.09)\\$[<0.005]$} &
	\shortstack{\textbf{52.07}\\($\pm$ 0.04)\\$[<0.005]$} \\
	\bottomrule
\end{tabular}%

		}
		\begin{tablenotes}
			\scriptsize
			\item[] Higher values are better. Parentheses show judge-run standard deviations.
			Brackets show Holm-adjusted two-sided Wilcoxon signed-rank $p$-values against Baseline 
       \item[] after Shapiro--Wilk rejected normality.
		\end{tablenotes}
	\end{threeparttable}
\end{table*}

\paragraph{Exploratory Model-Editing Comparison.}
\gls{taft} achieves the highest overall accuracy compared to GRACE~\cite{Hartvigsen2023GRACE}, MEMIT~\cite{meng2022mass}, and the non-finetuned base model, as shown in \cref{taft:tab:exploratory}.
Similarly, \gls{taft} also improves overall accuracy on the second model family Qwen2.5-Coder-14B, improving the overall accuracy from 38.34\% to 67.91\%.
The exploratory results suggest that \gls{taft} transfers to other model families and outperforms model editing approaches on continuous temporal adaptation.

\begin{table}[t]
	\renewcommand{\arraystretch}{0.8}
	\setlength{\tabcolsep}{2.5pt}
	\caption{Exploratory Accuracy Comparison}
	\label{taft:tab:exploratory}
	\begin{threeparttable}
		\resizebox{\columnwidth}{!}{
			\begin{tabular}{lcccc}
	\toprule
	\textbf{Approach} & \textbf{Change (\%) $\uparrow$} & \textbf{Retention (\%) $\uparrow$} & \textbf{Neighbor (\%) $\uparrow$} & \textbf{Overall  (\%) $\uparrow$} \\
	\midrule
	GPT-OSS-20B / Base model & 28.12          & 38.74          & 33.93          & 33.60 ($\pm$ 0.13)          \\
	GPT-OSS-20B / MEMIT      & 33.30          & 35.81          & 36.66          & 35.26 ($\pm$ 0.05)          \\
	GPT-OSS-20B / GRACE      & 24.55          & 41.56          & \textbf{42.02} & 36.04 ($\pm$ 0.01)          \\
	GPT-OSS-20B / TAFT       & \textbf{72.57} & \textbf{66.43} & 38.73          & \textbf{59.24} ($\pm$ 0.10) \\
	\midrule
	Qwen2.5-Coder-14B / Base model         & 34.99          & 41.86          & 38.16          & 38.34 ($\pm$ 0.01)          \\
   Qwen2.5-Coder-14B / TAFT               & \textbf{67.61} & \textbf{86.50} & \textbf{49.63} & \textbf{67.91} ($\pm$ 0.08) \\
	\bottomrule
\end{tabular}

		}
		\begin{tablenotes}
			\scriptsize
			\item[] Mean accuracies are calculated over three temporal slices and three runs.
			\\Parentheses indicate standard deviations.
		\end{tablenotes}
	\end{threeparttable}
\end{table}

\subsection[RQ2: Contribution of TAFT Components]{\rqb:~Contribution of \gls{taft} Components}
To isolate the contribution of each component, we conducted an ablation study comparing the \gls{taft} approach against two variants.
Specifically, we compare:
\begin{enumerate*}[label=(\roman*)]
	\item \gls{sft}-only, which observes only the currently valid answers $a^{+}$, and
	\item \gls{taft} without retention, which applies \gls{mpo} using temporal preference pairs $(a^{+}, a^{-})$
\end{enumerate*}

The ablation results show that the objectives contribute complementary behavior, with the full \gls{taft} approach achieving the best overall performance, as shown in \cref{taft:tab:ablation}.
Compared with \gls{sft}-only, the \gls{taft} variant without retention improves change accuracy from 71.01\% to 95.40\%, supporting the role of temporal preferences in overwriting stale answers.
However, both retention and neighbor accuracy decrease slightly compared to the \gls{sft}-only variant, from 28.39\% to 27.68\% and from 30.45\% to 29.42\%, respectively.
In contrast, the full \gls{taft} objective balances these effects, achieving 85.00\% change accuracy while substantially improving retention to 63.34\% and neighbor accuracy to 46.08\%.
Consequently, it achieves the highest overall accuracy of 64.81\%, compared with 50.83\% without retention and 43.28\% for \gls{sft}-only.
This indicates that \gls{mpo}-only updates can improve adaptation to changed examples, but they do so without preserving still-valid or semantically adjacent knowledge.
Thus, \gls{mpo}-only is insufficient, and the results highlight the need for the interaction between the components.
Overall, neither component is sufficient in isolation: preference learning provides temporal overwrite pressure to learn updated knowledge and suppress stale answers, while retention replay prevents degradation of still-valid knowledge.
Together, these findings support \gls{taft} as a fused temporal objective for repository-level code comprehension, where evolution invalidates some answers while leaving many related ones unchanged.

\begin{table}
	\caption{Mean Ablation Study Accuracies ($\pm$ Std.) over Three Runs.}
	\label{taft:tab:ablation}
	\resizebox{\columnwidth}{!}{
		\begin{tabular}{
		lcccc
	}

	\toprule

	\textbf{Approach}  & \textbf{Change (\%) $\uparrow$} & \textbf{Retention (\%) $\uparrow$} & \textbf{Neighbor (\%) $\uparrow$} & \textbf{Overall (\%) $\uparrow$ } \\
	\midrule
   SFT-only & 71.01 ($\pm$ 0.07) & 28.39 ($\pm$ 0.06) & 30.45 ($\pm$ 0.04) & 43.28 ($\pm$ 0.04) \\
   TAFT w/o retention & \textbf{95.40} ($\pm$ 0.02) & 27.68 ($\pm$ 0.02) & 29.42 ($\pm$ 0.09) & 50.83 ($\pm$ 0.04) \\
   TAFT & 85.00 ($\pm$ 0.04) & \textbf{63.34} ($\pm$ 0.04) & \textbf{46.08} ($\pm$ 0.02) & \textbf{64.81} ($\pm$ 0.03) \\
	\bottomrule
\end{tabular}

	}
\end{table}

\begin{custombox}[\rqb~ -- In summary]
	The ablation shows that the individual components are insufficient for robust temporal adaptation and confirms their complementary roles in \gls{taft}.
	\gls{mpo} improves adaptation to updated knowledge and suppression of outdated answers, but without retention it fails to preserve still-valid knowledge.
	Only the combined \gls{taft} objective provides the best overall performance across the evaluated dimensions.
\end{custombox}

\subsection[RQ3: Update Timing]{\rqc:~Update Timing}
RQ3 evaluates the cost--accuracy trade-off of continuous temporal adaptation.
\Cref{taft:tab:costs} reports the average accuracy, runtime, energy consumption, and carbon emissions across the evaluated approaches.
The results show a clear trade-off between adaptation accuracy and training cost.

The baseline has the lowest runtime, requiring 1.07h and 1.28 kWh while achieving 44.15\% overall accuracy.
Baseline-Batched does not improve accuracy, reaching 42.99\%, while increasing runtime to 6.99h and energy consumption to 7.30 kWh.
Similarly, augmenting the baseline with knowledge anchors does not improve accuracy or runtime, with Baseline-KANT reaching 35.31\% accuracy in 1.25h.

In contrast, \gls{taft} improves accuracy over all baseline variants, achieving 64.81\% compared with 44.15\% for the baseline, 42.99\% for Baseline-Batched, and 35.31\% for Baseline-KANT.
Compared to the baseline, this corresponds to a 46.8\% relative improvement in accuracy, at the cost of 2.58$\times$ higher runtime and 2.73$\times$ higher energy consumption.
Augmenting \gls{taft} with knowledge anchors~\cite{wolf.kant} further improves accuracy to 69.57\%, but also increases costs to 3.44h and 3.96 kWh, respectively.

Based on a Pareto-frontier analysis, the \gls{taft} variants provide strong cost--accuracy trade-offs, with \gls{taft} favoring lower computational cost and \gls{taft}-\gls{kant} trading additional cost for higher accuracy.

Overall, these results show that continuous temporal adaptation is preferable when repository freshness matters.
Both the baseline and Baseline-KANT variants are less costly but also less effective at preserving temporal alignment.
The \gls{taft} variants balance accuracy and adaptation cost by applying targeted local updates, trading higher training time and energy use for substantially improved alignment with repository evolution.

\begin{table}[hbt]
	\caption{Cost--Accuracy Trade-Off Across Approaches.}
	\label{taft:tab:costs}
	\begin{threeparttable}
		\resizebox{\columnwidth}{!}{
			\renewcommand{\arraystretch}{0.8}
\setlength{\tabcolsep}{3pt}
\begin{tabular}{lcccc}
	\toprule
	Approach         & \textbf{Accuracy (\%) $\uparrow$ } & \textbf{Time (h) $\downarrow$} & \textbf{Energy (kWh) $\downarrow$} & \textbf{CO$_2$ (kg) $\downarrow$} \\
	\midrule
	Baseline         & 44.15                              & \textbf{1.07}                  & 1.28                               & \textbf{0.49}                     \\
	Baseline-Batched & 42.99                              & 6.99                           & 7.30                               & 2.78                              \\
	Baseline-KANT    & 35.31                              & 1.25                           & \textbf{1.25}                      & 0.56                              \\
	TAFT             & 64.81                              & 2.76                           & 3.50                               & 1.33                              \\
	TAFT-KANT        & \textbf{69.57}                     & 3.44                           & 3.96                               & 1.51                              \\
	\bottomrule
\end{tabular}

		}
		\begin{tablenotes}
			\scriptsize
			\item[] Accuracy reports the mean across dimensions, time slices, and runs.
			\item[] Energy metrics report the accumulated training cost across time slices.
		\end{tablenotes}
	\end{threeparttable}
\end{table}

\begin{custombox}[\rqc~ -- In summary]
	\gls{taft} variants provide strong cost--accuracy trade-offs, substantially improving accuracy over the baselines while maintaining continuous temporal freshness.
	These results show that continuous temporal adaptation is preferable when models must remain aligned with repository evolution.
\end{custombox}

\subsection[RQ4: Locality and Stability]{\rqd:~Adaptation Locality and Stability}
RQ4 investigates whether temporal updates remain localized to changed knowledge or spill over into semantically neighboring but unaffected knowledge.
The neighbor accuracy results show that \gls{taft} consistently outperforms the baseline.
Across the temporal stream, \gls{taft} reaches 46.08\% average neighbor accuracy compared to 31.19\% for the baseline, corresponding to approximately \num{1.48}$\times$ the baseline performance.

The gain is consistent across all ten time slices, indicating that \gls{taft} reduces unintended effects on semantically related but unchanged knowledge.
Augmenting the baseline with knowledge anchors does not improve neighbor accuracy. Baseline-\gls{kant} decreases average neighbor accuracy from 31.19\% to 29.98\%.
In contrast, \gls{taft}-\gls{kant} further increases neighbor accuracy from 46.08\% to 52.07\%, corresponding to a relative improvement of 13.00\%, and outperforms \gls{taft} across all ten temporal slices.
This pattern suggests that knowledge anchors may provide additional cues for distinguishing changed and suppressed knowledge from retained and neighboring knowledge within the \gls{taft} update objective.
This behavior is consistent with the role of retention examples, which constrain broad shifts while \gls{mpo} updates changed behavior.
Still, neighbor accuracy remains the lowest of the evaluated dimensions.
A likely reason is that neighbor questions are intentionally close to updated questions, often sharing context and identifiers, making it difficult for the model to distinguish updated from related knowledge.
Conversely, neighbors that overlap more strongly with retained \gls{qa} pairs may be less affected, suggesting that retention partly mitigates, but does not eliminate, collateral updates.
The reported neighbor accuracy should therefore be interpreted as relative correctness on adjacent unchanged knowledge, not as a complete explanation of stepwise collateral damage.
This remaining gap indicates that preserving nearby unaffected knowledge is still an open challenge and motivates future work on stronger locality constraints, more fine-grained neighbor sampling, or explicit regularization over semantically adjacent but unchanged examples.

\begin{custombox}[\rqd~ -- In summary]
	\gls{taft} consistently improves accuracy on neighbor knowledge across all temporal slices, increasing the average accuracy by $\approx 1.48\times$ over the baseline.
	TAFT augmented with knowledge anchors further increases neighbor accuracy from 46.08\% to 52.07\%.
	However, neigbhor accuracy remains the weakest dimension, indicating that retention reduces collateral updates but does not fully prevent overgeneralization to semantically adjacent knowledge; preserving such knowledge remains an open challenge that motivates further analysis.
\end{custombox}

\section{Threats to Validity}\label{taft:sec:threats}
\textit{Construct validity.}
We operationalize adaptation through generated \gls{qa} tuples for updated, outdated, retained, and neighboring knowledge, exposing the temporal relation between $t-1$ and $t$.
However, this abstraction cannot capture all comprehension needs, such as multi-file reasoning, architectural intent, runtime behavior, or external documentation.
Moreover, the boundary between changed, retained, and neighboring knowledge can be blurred by partial invalidation or indirect cross-file effects.
We mitigate this risk by grounding generation and verification in the \gls{pr} description, patch, target file context, and paired current and outdated answers.

\textit{Internal validity.}
Internal validity concerns whether observed differences can be attributed to \gls{taft} rather than confounding design choices.
Our ablations keep the benchmark splits, model initialization, configurations, and evaluation protocol fixed, supporting controlled comparison between the components and their combination.
Training and evaluation examples originate from the same repository transitions~\cite{kaufman2012leakageindm}, and retention examples contribute to the training objective~\cite{rolnick2019experience,tiwari2022CVPR}.
This overlap is intentional, as we evaluate supervised knowledge maintenance across successive updates rather than generalization to unseen questions.
Stochastic training, retention sampling, hyperparameters, and temporal ordering can also affect fine-tuning outcomes~\cite{cha2025hyperparameters}.
Thus, the results support the proposed objective under the stated protocol, but not under all training budgets, sampling policies, or update schedules.
Baseline-KANT and TAFT-KANT are evaluated with directly supplied correct anchors and without the \gls{fim} stage of the complete \gls{kant} pipeline.
Thus, their results do not measure automatic anchor-selection or the performance of the full \gls{kant} training procedure.
The exploratory comparison introduces additional confounding factors as neither GRACE~\cite{Hartvigsen2023GRACE} nor MEMIT~\cite{meng2022mass} was designed for \gls{moe} architectures.
Thus, the observed performances may reflect architecture-specific constraints rather than effectiveness under their intended settings.

The study relies primarily on \gls{llm}-based judging of \gls{qa} pairs.
This scales our dataset, but may introduce biases: generated questions can omit implicit context, stay too close to patches, or favor facts that are easy to express.
The verification prompt removes pairs that are unsupported by the code evidence, but cannot guarantee correctness like expert annotation or executable checks.
Similarly, the \gls{llm}-judge may favor reference-like phrasing, miss subtle outdated knowledge, or vary across question types.
We mitigate these risks by validating the judge against adjudicated human labels on a stratified sample of \num{400} items, using the same judging protocol for all systems, and evaluating all dimensions separately across three runs.
The human validation does not provide labels for the full dataset. Therefore, absolute accuracy values should still be interpreted cautiously.

\textit{External validity.}
Our benchmark builds on SWE-bench, which provides realistic, test-validated GitHub issue resolutions across popular Python repositories.
However, it emphasizes issue-fixing patches and may not capture refactoring, other changes, non-Python repositories.
We therefore view the results as evidence for temporal adaptation in this controlled Python setting, with cross-language replications left for future work.

\textit{Temporal validity.}
Our temporal stream is constructed from version tags and normalized repository-specific bins.
This makes heterogeneous histories comparable and reproducible, but only approximates how repository knowledge changes in deployment.
Tags can be sparse, inconsistent, or delayed relative to the commits that introduced a change.
Normalizing histories by relative position also removes absolute calendar time and release cadence, both of which may affect update frequency decisions.
Thus, the update-timing analysis captures quality--cost trade-offs under our discretized stream, not a universal prescription for retraining in every workflow.

\section{Related Work}\label{taft:sec:relatedWork}
The work aims to equip repository-specialized \glspl{llm} with the ability to remain temporally consistent as software evolves.

\paragraph{Repository-level code comprehension, QA, and datasets.}
Prior work on developer information needs shows that developers ask questions about behavior, rationale, and usage relationships, often requiring evidence from multiple artifacts~\cite{ko2007information,latoza2010hard}.
Recent \gls{llm}-based work extends these needs to repository-scale settings, including in-IDE code understanding and benchmarks for documentation, and code generation~\cite{nam2024using,luo2024repoagent,strich2024repositoryqa,wolf.kant}.
These efforts build on datasets that serve as important precursors by exposing models to realistic code context through related tasks, such as code search, repair, and generation~\cite{jimenez2024swebench,liu2021CodeQA,liu2023repobench,ding2023crosscodeeval}.
Nevertheless, these datasets typically represent a task, repository snapshot, or retrieval/completion problem, and prior work treats repository comprehension as single-snapshot learning focused on acquiring new knowledge.
In contrast, \gls{taft} uses repository transitions, deriving updated, outdated, retained, and neighbor \gls{qa} examples to assess adaptation quality across multiple dimensions.

\paragraph{Software evolution and stale model knowledge.}
Software-evolution research treats repository knowledge as temporally unstable.
Studies of documentation, comments, dependencies, and library migrations show that artifacts often lag behind changes~\cite{rani2023CommentQuality,tan_wait_2023,he_large-scale_2021}.
For code \glspl{llm}, recent work shows that models emit deprecated APIs under library evolution and explores version-aware API migration~\cite{LLMDeprecation2024,ReCode2025}.
Recent benchmarks similarly emphasize data freshness under code evolution~\cite{Lin2025RustEvo,li2024evocodebench}.
Together, these efforts identify stale model knowledge as a practical problem, but focus on code generation, completion, or API migration.
\gls{taft} instead targets repository-comprehension \gls{qa} for evolving repositories, where adaptation must prefer the answer valid at $r_t$, reject the answer valid only at $r_{t-1}$, and preserve answers unaffected by the transition.

\paragraph{Continual adaptation, suppression, and retention.}
Continual learning studies how models acquire new tasks or distributions without catastrophic forgetting~\cite{mccloskey_catastrophic_1989}.
In code, Yadav et al.~\cite{yadav2023exploring} instantiate this setting through CodeTask-CL, where models adapt sequentially across tasks and use replay to retain performance.
However, conventional learning episodes are usually defined by tasks, domains, or concepts, whereas \gls{taft} defines them by concrete transitions between repository states.
This granularity is essential because repository evolution separates newly valid, obsolete, and still-valid behavior, a distinction conventional continual-learning formulations do not encode directly.

Replay is a principal mechanism for stabilizing learning, revisiting prior examples while learning new data~\cite{rolnick2019experience,lange_continual_2021}.
In \glspl{llm}, rehearsal can use retained data, synthetic replay, distillation, or attention-guided retention to mitigate forgetting~\cite{huang2024mitigating,he2024seekr}.
\gls{taft} adopts this intuition but constrains replay to repository evolution: retention examples are not arbitrary samples, but facts that remain valid across the same transition that introduces changed and outdated answers.
Replay alone is insufficient, since it preserves prior behavior but does not identify which prior answers should be suppressed.
\gls{taft} therefore couples retention replay with temporal preference supervision over paired $(a^{+},a^{-})$ answers.

Suppressing obsolete behavior is also related to \gls{mu} and model-editing.
\Gls{mu} removes selected data or behaviors, often for privacy or deletion requirements~\cite{chen_unlearn_2023,liu2025rethinking}, while model-editing modifies targeted factual associations through parameter edits, external memories, or mass-editing mechanisms~\cite{meng2022locating,meng2022mass,Hartvigsen2023GRACE}.
However, repeated edits can weaken previously edited knowledge and cause gradual or catastrophic forgetting~\cite{li2024continualedit,gupta2024modelscale}, and editing code models remains difficult in generalization settings~\cite{Li2025ModelEditingLLMs4Code}.
Nevertheless, their assumptions differ from repository adaptation: unlearning treats information as removable, and editing treats target associations as patchable, whereas repository evolution requires suppressing only obsolete behavior while preserving neighboring and historical knowledge that remains valid.

\Gls{sft} and preference optimization are common mechanisms for aligning language models with task-specific goals.
\gls{dpo} optimizes preferences between accepted and rejected responses, while related objectives combine preference and likelihood signals in different ways (e.g. ORPO and MPO)~\cite{rafailov2023dpo,hong2024orpo,wang2024mpo}.
These methods combine pairwise and likelihood-based training, but their labels usually encode alignment or response quality.
In \gls{taft}, the preference is temporal rather than qualitative: repository evolution determines that $a^{+}$ is current and $a^{-}$ is stale.
Similarly, the supervised component is not a general quality regularizer, but a replay term over answers that should remain correct after the code change.
Thus, \gls{taft} combines suppression and retention into a single temporal adaptation objective grounded in code evolution.

\section{Future Work}\label{taft:sec:futureWork}
A future direction for \gls{taft} is broader validation across languages as our current benchmark builds on SWE-bench, which focuses on Python.
Extending \gls{taft} to languages such as JavaScript, C++, and Java would clarify the generalizability of \gls{taft} across languages.%

The most important methodological extension concerns neighboring knowledge.
In the current design, neighbor examples are used to measure locality and stability rather than as direct update constraints,
and the results show that neighbor stability remains the weakest dimension.
Future work should therefore examine how neighbors are defined, sampled, and represented, as well as how they can be incorporated into the training objective.
Promising directions include version-aware or finer-grained knowledge anchors to distinguish changed, retained, and neighboring concepts, as well as neighbor-aware regularization, contrastive locality losses, and consistency constraints.
Such objectives would make locality an explicit optimization target rather than just an evaluation criterion.
The anchor-conditioned variants should also be evaluated with automatically selected anchors used in the complete \gls{kant} pipeline~\cite{wolf.kant}.

Another direction is adaptive update scheduling.
This paper uses normalized chronological bins for controlled comparison.
In practice, updates may be triggered by commit density, release boundaries, dependency changes, API churn, or measured model drift.
Future work should compare adaptive update schedules with Baseline-Batched and the other evaluated update strategies, measuring the trade-off between freshness, knowledge preservation, and training cost.

Finally, the human validation provides an independent assessment of the \gls{llm}-judge on a stratified sample, while the full-scale evaluation remains automated.
Future work should extend this validation with domain-expert annotation, repository tests, and developer studies.

\section{Conclusion}\label{taft:sec:conclusion}
This work aims at improving repository-specialized \glspl{llm} with the ability to remain temporally consistent as software evolves,
aligning models alongside the evolution by adapting to new knowledge, suppressing obsolete answers, retaining stable knowledge,
and preserving neighboring semantics.
\gls{taft} addresses this by treating transitions as a temporal preference between updated and stale knowledge, while
retaining stable knowledge through transition-conditioned retention replay.

Our evaluation shows that \gls{taft} improves each evaluated dimension of temporal repository adaptation.
For updated and suppressed knowledge, \gls{taft} significantly improves over the baseline by more than \num{18}\% (p-value $\leq 0.005$),
demonstrating \gls{taft}'s ability to align models to the updated knowledge, while suppressing obsolete knowledge through temporal preference supervision.
For retention, \gls{taft} achieves an average accuracy of more than double that of the baseline,
showing that transition-conditioned replay substantially reduces forgetting of stable knowledge.
For neighboring knowledge, \gls{taft} reaches \num{46.08}\% average accuracy compared to \num{31.19}\% for the baseline, a \num{1.48}$\times$ improvement while also showing that neighboring semantics remain an open challenge.
Under directly supplied anchors, \gls{taft}-\gls{kant} further increases neighbor accuracy to \num{52.07}\%.
Furthermore, the \gls{taft} variants provide strong cost--accuracy trade-offs.
It is more costly than the baseline, requiring \num{2.76}h of training time compared to \num{1.07}h, but this added cost enables
substantially higher accuracy and continuous temporal freshness.
Thus, showing that explicit temporal adaptation yields better adaptation quality at the cost of additional training time and energy.

The ablation study further explains where these gains come from.
It confirms that preference learning alone is insufficient, and that the strongest results require the full objective combining updated, suppressed, and retained knowledge.
Thus, \gls{taft}'s effectiveness depends not only on learning what changed, but also on jointly suppressing what expired and replaying what must remain valid.

These findings advance repository-level \gls{llm} adaptation, by making software evolution trainable: it distinguishes knowledge that changed, knowledge that expired, and knowledge that must be preserved.
This gives repository-specialized models a practical path to remain aligned with evolving code without being rebuilt from scratch.

The approach remains bounded by the constructed benchmark, backbone model family, and \gls{llm}-judge protocol.
Future work should strengthen locality, study adaptive update schedules, and replicate the results across more languages, models, and repository types.

Overall, \gls{taft} demonstrates that repository-specialized \glspl{llm} can be adapted along code evolution by learning what changed, retaining what remains valid, and suppressing what became obsolete.
This makes temporal alignment a concrete and measurable requirement for future code comprehension systems.

\section*{Acknowledgments}

The research leading to these results has been supported by the Swiss National Science Foundation (SNSF) through Grant SNSF204632.

\bibliography{reference}
\end{document}